\documentclass[aps,pra,twocolumn,amssymb,amsfonts,floatfix,longbibliography]{revtex4}
\usepackage{bm}
\usepackage{dcolumn}
\usepackage{bm}
\usepackage{mathbbol}
\usepackage{graphicx}
\usepackage{epstopdf} 
\usepackage{amsmath,amsthm} 
\usepackage{times}
\usepackage{color}
\usepackage{braket}
\usepackage{placeins}
\FloatBarrier

\begin{document}

\title{Disorder enhanced transport as a general feature of long-range hopping models
}
\author{Elisa Zanardini}
\affiliation{Dipartimento di Matematica e Fisica and ILAMP, 
Università Cattolica del Sacro Cuore, Brescia, Italy}
\affiliation{Istituto Nazionale di Fisica Nucleare, sez. Milano, Milano, Italy}
\affiliation{Department of Physics and Astronomy, University of Notre Dame, Notre Dame, Indiana 46556, USA}
\author{Giuseppe Luca Celardo}
\affiliation{Department of Physics and Astronomy and CSDC, University of Florence, Florence, Italy.}
\affiliation{European Laboratory for Non-Linear Spectroscopy (LENS), University of Florence, Florence, Italy, 50019 Sesto Fiorentino, Italy}
\affiliation{Istituto Nazionale di Fisica Nucleare (INFN), Sezione di Firenze, 50019 Sesto Fiorentino, Italy.}
\author{Nahum C. Ch\'avez}
\affiliation{Dipartimento di Matematica e Fisica and ILAMP, 
Università Cattolica del Sacro Cuore, Brescia, Italy}
\author{Fausto Borgonovi}
\affiliation{Dipartimento di Matematica e Fisica and ILAMP, Università Cattolica, Brescia, Italy}
\affiliation{Istituto Nazionale di Fisica Nucleare,  sez. Milano, Milano, Italy}

\date{\today}

\begin{abstract}
We analyze the interplay of disorder and long-range hopping in a paradigmatic  one dimensional model of quantum transport.  While typically the current is expected to decrease as the disorder strength increases due to localization effects, in systems with infinite range hopping it was shown in  Chavez et al, Phys. Rev. Lett. {\bf 126}, 153201 (2021), that the current can increase with disorder in the Disorder-Enhanced-Transport (DET) regime. 
Here, by analyzing  models with variable hopping range decaying as $1/r^{\alpha}$ with the distance $r$ among the sites, we show that the DET regime is a general feature of long-range hopping systems and it occurs, not only in the strong long-range limit $\alpha<1$ but even for weak long-range $1 \le \alpha \le 3$. Specifically, we show that, after an initial decrease, the current grows with the disorder strength  until it reaches a local maximum. Both  disorder thresholds at which the DET regime starts and ends are determined.  Our results open the path to understand the effect of disorder on transport in many realistic systems where long range hopping is present.
\end{abstract}
\maketitle

\section{Introduction}

Understanding quantum transport in disordered systems remains a central problem in condensed matter physics, with implications for both fundamental science and applied quantum technologies. In one-dimensional (1D) systems with nearest-neighbor hopping, where each site can interact only with its neighbors, disorder typically leads to Anderson localization, suppressing transport exponentially with the system size and disorder strength \cite{anderson1958absence}. While Anderson localization has been a beacon in the field of quantum transport, its validity relies on some assumptions, for instance that the hopping is nearest-neighbor. Extending our understanding of the intrinsic response of a quantum system to disorder beyond the Anderson localization paradigm is essential to understand the behavior of many realistic systems. 

Recent work has shown the existence of novel transport regimes in the presence of distance independent long-range hopping~\cite{chavez2021disorder}. In particular, it was shown in Ref.~\cite{chavez2021disorder} that for a fully connected 1D chain with constant long-range hopping, the interplay between long-range hopping and disorder leads to the emergence of two distinct transport regimes. In the \textit{Disorder-Enhanced Transport} (DET) regime, the current increases with the strength of the disorder, while in the \textit{Disorder-Independent Transport} (DIT) regime, the current remains unaffected by the increase in disorder over several orders of magnitude of disorder strength. 
These results have been discussed in several recent works for various models, both theoretically ~\cite{allard2022disorder, engelhardt2023polariton, wang2023quantum, dominguez2021enhanced, wu2024disorder, wellnitz2022disorder, engelhardt2022unusual, aroeira2024coherent, krupp2025, suyabatmaz2023vibrational, gera2022effects, sun2022dynamics, ciuti2021cavity, coates2023goldilocks, pouranvari2024probing, davidson2022eliminating, zhou2024nature} and experimentally ~\cite{george2024controlling, baghdad2023spectral, sandik2025cavity, cohn2022vibrational, hamerlynck2022static, toffoletti2025coherent, balasubrahmaniyam2023enhanced, sauerwein2023engineering}. 
Note that the DET regime found in presence of distance-independent long-range hopping shares some resemblance with the enhanced-noise-assisted-transport regime (ENAQT)~\cite{rebentrost2009environment, plenio2008dephasing, caruso2009highly}. Nevertheless, the ENAQT regime occurs when the intrinsic transport is very weak due to interference effects so that disorder or noise can enhance it. On the other side, here the  claim is that the DET regime is an intrinsic property of long-range hopping systems and it exists even if transport is intrinsically efficient.  

The DET and DIT regimes in presence of all to all hopping, had been explained in terms of the hybrid nature of the excited eigenstates which are characterized by Anderson-like localized peaks and extended flat tails.
 Extended plateaus allow transport, thus overcoming the standard view of the transport suppression   due the exponential localization of the eigenstates. These findings set a new paradigm for transport in presence of disorder and long-range hopping. 
Moreover, these results are not only of theoretical interest: effective long-range couplings can be emulated in ion-traps~\cite{jurcevic2014quasiparticle} and are relevant in several realistic systems such as cold atomic clouds~\cite{guerin2017light} and excitonic transport in molecular aggregates~\cite{struempfer2012quantum, gulli2019macroscopic, spano1991cooperative}.\\

In this work, we generalize the scenario of Ref.~\cite{chavez2021disorder} by introducing a distance-dependent long-range hopping of the form $-\gamma/r^\alpha$, where $\alpha$ tunes the hopping range. This allows us to interpolate between the all-to-all model ($\alpha = 0$) and the nearest-neighbor limit ($\alpha \rightarrow \infty$) and to explore how transport regimes evolve under varying degrees of locality. Unlike previous studies, our model includes only power-law hopping, omitting nearest-neighbor terms, and thus isolates the role of interaction range in shaping transport behavior.\\
Using a Lindblad master equation formalism to model pumping and draining at the chain ends of a one dimensional quantum model, we compute the steady-state current and analyze its properties. 
While the DIT regime in 1D systems arises only for distance-independent hopping, here we show that the DET regime is a universal feature of long-range hopping and it persists, not only for strong long range hopping ($\alpha<1$) but even for weak long range hopping ($1 \le \alpha \le 3$). 
Following the procedure adopted in Ref.~\cite{chavez2021disorder} we further derive novel  expressions for the critical disorder thresholds that mark the onset and breakdown of the DET regime.\\
\section{The Model}
\begin{figure}
\centering
\includegraphics[width=3in]{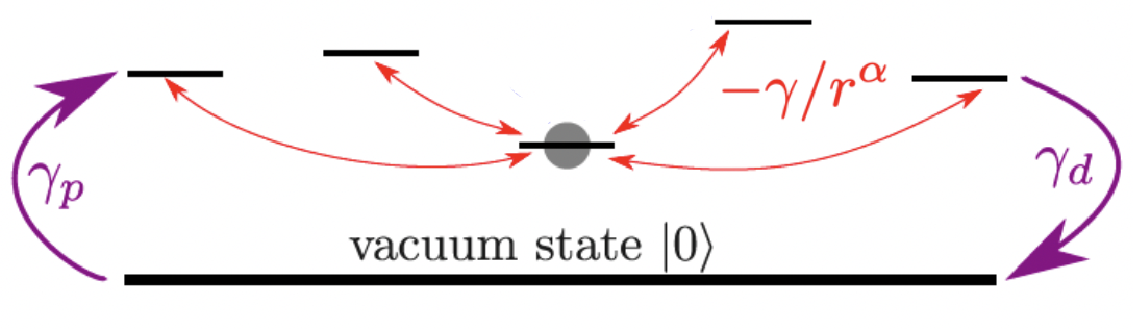}
\caption{A disordered chain with excitation pumping $\gamma_p$ at one edge of the chain and draining $\gamma_d$ at the opposite edge. The energy of the sites is disordered. Here, $-\gamma/r^\alpha$ is the distance-dependent long-range hopping between each pair of sites.}
\label{fig:fig1}
\end{figure}

As a paradigmatic model of a disordered chain in presence of a distance-dependent long-range hopping, we analyze a linear-chain with hopping amplitude $\gamma$ among sites $\ket{i}$ staggered by a random local potential $\epsilon_i$, according to the Hamiltonian:
\begin{equation}
H_\alpha = H_0 + V_\alpha = \sum\limits_{i=1}^N {\epsilon}_i |i\rangle\langle i| -\gamma \sum_{j\neq i} {|i\rangle \langle j|\over{|i-j|^{\alpha}}}
\label{ham}
\end{equation}
The coefficient $\alpha$ defines the hopping range. According to standard literature, we define the regime of long-range hopping when $\alpha \leq d$, and $\alpha > d$ for short-range hopping, where $d$ is the dimension of the system (here $d=1$). The case $\alpha = 0$ corresponds to the all-to-all long-range hopping (infinite range) while by setting $\alpha = \infty$ we retrieve the Anderson model with an effective nearest-neighbor hopping $- \gamma$. Let us note that $\alpha = 1$ may be considered a critical case, since the interaction range   match the system dimension $d=1$~\cite{celardo2016shielding}. 
To understand how transport properties are affected by a finite long-range hopping, we analyze the stationary current as a figure of merit of transport efficiency, expanding the results presented in \cite{chavez2021disorder} for the specific case $\alpha = 0$ to the more general case $\alpha > 0$.  In order to compute the current, we introduce pumping and draining at the chain edges, as shown in the system representation in Fig. \ref{fig:fig1}.
The dynamics is then described by the Lindblad Master Equation \cite{breuer_theory_2002}:
\begin{equation}
\frac{d\rho}{dt} = -\frac{i}{\hbar}[H, \rho] + \sum_{\eta = p,d} \mathcal{L}_\eta [\rho],
\label{Lindblad}
\end{equation}
where $\mathcal{L}_\eta [\rho] = -\{L_\eta^\dagger L_\eta\, \rho\} + 2L_\eta \rho L_\eta^\dagger
$ with $\eta=p,d$ are two dissipators inducing pumping on the first site ($L_p = \sqrt{\gamma_p/(2\hbar)} \ket{1}\bra{0}$) and draining at the last site ($L_d = \sqrt{\gamma_d /(2\hbar)} \ket{0} \bra{N}$) respectively, and $\ket{0}$ represents the vacuum state. From the steady-state solution of Eq. (\ref{Lindblad}) one can find the stationary current
\begin{equation}
I = \frac{\gamma_d}{\hbar} \bra{N} \rho_{SS} \ket{N}
\end{equation}
where $\rho_{SS}$ is the steady-state density operator. \\
It is possible compute the current using a non-Hermitian Schr\"odinger equation as proposed in~\cite{chavez2021disorder}. This method gives a current which is  exactly equivalent to the current computed with the Master Equation, as demonstrated in~\cite{chavez2021disorder}. In particular, it is possible to compute the probability amplitude on the drain site $\psi_N(t)$ at time $t$ under the effective Hamiltonian \cite{celardo2012superradiance, zhang2017opening},
\begin{equation}
(H_{eff})_{k,l} = (H)_{k,l} -i{\gamma_d\over{2}}\delta_{k,N}\delta_{l,N}
\label{effham}
\end{equation}
and use it to compute the average transfer time $\tau $ \cite{kropf2019towards, kropf2019electric, mohseni2008environment, rebentrost2009environment}. The latter is defined as the average time needed for excitations to leave the 1D chain:
\begin{equation}
\tau = {\gamma_d\over{\hbar}}\int_0^{\infty}dt\,t |\psi_N(t)|^2.
\end{equation}
This integral can be evaluated analytically\cite{chavez2021disorder} by expanding $e^{-iH_{eff}t}$ on the eigenbasis of $H_{eff}$. Being $H_{eff}$ non-Hermitian, it has right and left eigenvectors:
\begin{equation}
H_{eff}|r_k\rangle = \epsilon_k|r_k\rangle
    \; \; \; \; \; \; \; \; \; \; \langle \tilde{r}_k|H_{eff} = \langle \tilde{r}_k| \epsilon_k
\end{equation}
Specifically, it can be shown \cite{chavez2021disorder} that   if the excitation is initially  on site $|1\rangle$, in presence of a drain in the last site $|N\rangle$    with draining rate $\gamma_d$, the average transfer time is given by: 
\begin{equation}
\tau = {\hbar\gamma_d \sum_{k,k'}{{{{\langle N |r_k \rangle \langle \tilde{r}_k | 1 \rangle \langle N|r_{k'} {\rangle}^{*}} \langle \tilde{r}_{k'} | 1 {\rangle}^{*}}}\over{-{(\epsilon_k - {\epsilon}_{k'}^{*})}^2}}}.
\label{tau}
\end{equation}
This equation is quite relevant since it shows that the average time depends not only on the energy difference between different complex energies of the non-hermitian Hamiltonian but also on the overlap of the eigenfunctions with the initial (pump) and the last (drain) site. The latter point is remarkable since it means that the average transfer time is independent of the probability amplitude on all the sites of the chain but the first and the last one. 

The steady-state population   can be derived, by assigning a drain frequency $1/\tau$ and a pumping frequency $\gamma_p/\hbar$ connecting the chain population $P_e$ to the vacuum state $\ket{0}$ with population $P_0$ \cite{chavez2021disorder}:
\begin{equation}
\frac{dP_0}{dt} = -\frac{\gamma_p}{\hbar}P_0 + \frac{1}{\tau} P_e
\end{equation}
with the condition:
\begin{equation}
P_0+P_e = 1
\end{equation}
The steady-state population $P_e^{SS}$ can then be used to compute the steady-state current:
\begin{equation}
I = \frac{P_e^{SS}}{\tau} = {1\over{\tau}}{{\gamma_p}\over{\gamma_p + {\hbar\over{\tau}}}}.
\end{equation}
 The average current is then computed over different realizations of the site energies $\epsilon_i$ randomly chosen among the interval $[-{W\over{2}}, {W\over{2}}]$. Thus $W$ defines the strength of the disorder.
The typical current $I^{typ}$ can be finally obtained as:
\begin{equation}
I^{typ} = e^{\langle lnI \rangle}.
\end{equation}
The usage of the typical current instead of the standard average current finds a justification in the fact that the first quantity has the self-averaging property. On the contrary, we checked that here, similarly to the results found in Ref.~\cite{chavez2021disorder}, the average current is not  self-averaging, meaning that it could not be possible to find a good estimate for the current by averaging over a sufficiently large sample. This lack of self-averaging originates from the fact that, as we will show in Section IIIc, in the relevant disorder range, the current is dominated by a very small subset of eigenstates. In particular, a single eigenstate often gives the largest contribution to transfer time. This produces a broad, heavy-tailed distribution of currents in which rare configurations dominate the average, preventing it from converging.\\
 \begin{figure*}[ht]
\centering
\includegraphics[width=\textwidth]{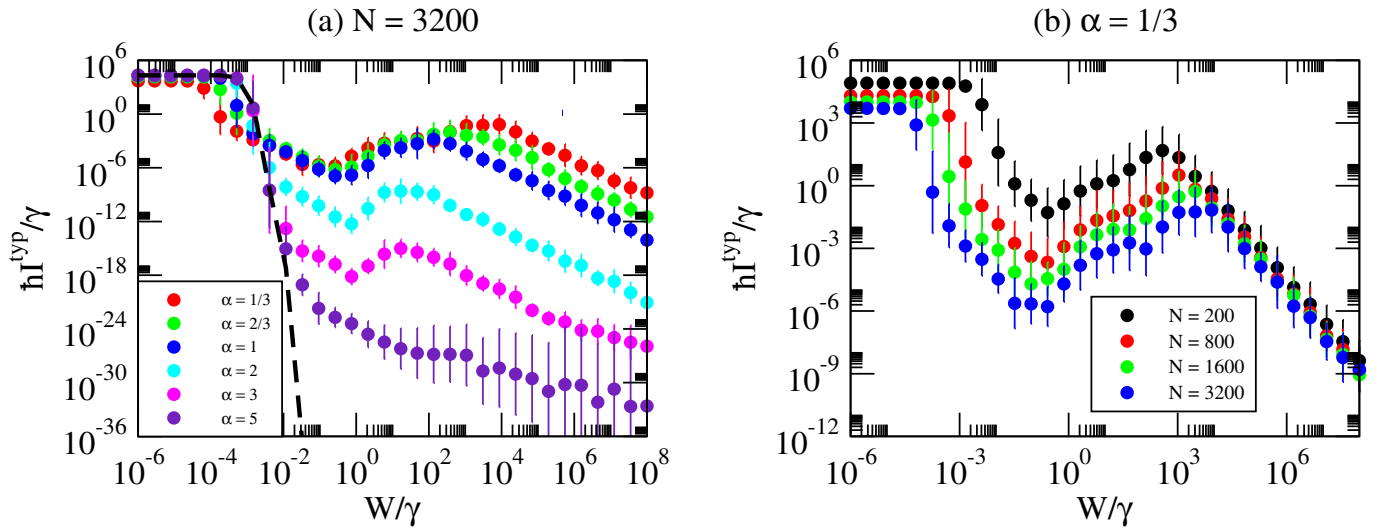}
\caption{
(a) Rescaled typical current $\hbar I^{typ}/\gamma$ as a function of disorder strength $W/\gamma$ for a chain of $N = 3200$ sites and several finite range interaction parameters $\alpha$. The black dashed line shows the case $\alpha = \infty$.
(b) Rescaled typical current $\hbar I^{typ}/\gamma$ as a function of disorder strength $W/\gamma$ for a long-range finite interaction range $\alpha = 1/3$ and different system sizes $N$. In both the panels the error bars are given by the standard deviation of the average over the different configurations and  $\gamma_p = \gamma_d = \gamma$ and $N \times N_r = 10^5$.}
\label{fig:newfig}
\end{figure*}
As already stated in the Introduction, the main result found in~\cite{chavez2021disorder}  was the presence of two distinct transport regimes called Disorder Enhanced Transport (DET) and Disorder Independent Transport (DIT) regimes. 
While the first regime is characterized by a current which, quite counter-intuitively, increases with the increase of the strength of disorder $W$, the latter regime is characterized by a large plateau in which the current does not depend on the disorder strength.

In Ref.~\cite{chavez2021disorder} a model with infinite hopping range $\alpha=0$ was considered with the same Hamiltonian as in Eq.~(\ref{ham}) with the addition of a nearest-neighbor hopping $\Omega$. The critical disorder strengths for the onset of the two regimes were determined  through an hand-waving argument. On  increasing the disorder strength the current decreases, as in a standard Anderson chain, but only    up to a critical   threshold, given by,
\begin{equation}
    \label{eq:DET}
W_1 = \Omega \sqrt{\frac{210.4 \ln N}{N}},  
\end{equation}
where $\Omega$ is the nearest-neighbor hopping. This disorder threshold has been found by imposing that the exponential peak at the chain edges $\exp (-N/(2\xi))$, where $\xi = 105.2(\frac{\Omega}{W})^2$ is the Anderson localization length, is almost equal to the probability in the tails $1/N$ \cite{chavez2021disorder}. After this disorder threshold, the current {\it increases} with disorder up to another disorder threshold found by imposing that the exponential peak at the closest site $\exp (-1/(2\xi))$ is almost equal to $1/N$ \cite{chavez2021disorder}:
\begin{equation}
    \label{eq:DIT}
W_2 = \Omega \sqrt{210.4 \ln N},  
\end{equation}
This second  regime (DIT) persists until the disorder strength becomes comparable with the energy gap between the ground state and the first excited state. The disorder strength at which the gap closes is given by:
\begin{equation}
    \label{eq:GAP}
W_{gap} = \frac{1}{2}\gamma N \ln N.  
\end{equation}
A further increase of the disorder strength produces a decrease of the current but with features different from the standard Anderson  picture (e.g. the currents decreases as a power law in $W$ instead of an exponential function). Note that the presence of an energy gap between the ground state and the first excited state is a general feature of long-range interactions. In particular, in the fully connected case, the spectrum can be computed analytically resulting in a fully symmetric ground state with energy $-\frac{\gamma}{2}(N-1)$ and excited states forming an $(N-1)$-fold degenerate manifold at energy $+\frac{\gamma}{2}$, yielding a gap $\Delta = \frac{\gamma N}{2}$ \cite{chavez2021disorder}. For general power-law hopping, the persistence of a finite gap is established in Ref. \cite{celardo2016shielding}. In particular, the eigenvalues are given by:
\begin{equation}
E_q^{LR}=\left\{ 
    \begin{split}
        -2\gamma \sum_{n=1}^{N/2-1} \frac{\cos{2\pi q n/N}}{n^\alpha} \qquad  N \quad  \text{Odd}\\
        -\gamma \frac{(-1)^q}{(N/2)^\alpha}-2\gamma \sum_{n=1}^{N/2-1} \frac{\cos{2\pi q n/N}}{n^\alpha} \qquad  N \quad \text{Even}
    \end{split}\right.
\end{equation}
and thus the energy gap scales as:
\begin{equation}
\Delta^{LR}\propto\left\{ 
    \begin{split}
        \gamma N^{1-\alpha} \quad\text{for}\quad \alpha \leq3\\
        \gamma N^{-2} \quad\text{for}\quad \alpha>3
    \end{split}\right.
\end{equation}

One should note that for large system size $N\to \infty$ $W_1\to 0$, while both $W_2, W_{gap} \to \infty$, even if $W_{gap} $ grows to $\infty$   much faster than $W_2$. This means that for any choice of the parameters $\gamma, \Omega $, and a  sufficiently large value of $N$   both DET and DIT regimes are well defined.

 From now on, all quantities related to energy (including energy $E$, energy gap $\Delta$, and disorder $W$) will be given in units of the long-range coupling strength $\gamma$.
 
We investigated whether the DET and DIT regimes persist when the hopping decays with distance. Specifically, we determine new characteristic disorder thresholds i.e. their dependence on the hopping range $\alpha$.\\
The main picture is shown in Fig.~\ref{fig:newfig} where the current is plotted {\it vs} the disorder strength $W$ for different values of $\alpha$ and fixed $N=3200$  in panel (a), while in panel (b) we show different $N$ values for a fixed  long-range hopping $\alpha=1/3$.  As one can see in panel (a) the current decays with disorder and then for   $\alpha \leq 3 $ it starts to increase up to a local maximum, after which it decays again with disorder. This shows that the DET regime is a general feature of the strong   ($\alpha < 1$)  and the  weak long-range hopping ($1\leq\alpha < 3$).  On the other side, the DIT regime does not occur  for any non zero hopping range $\alpha$. 
 Results from panels (a) can be summarized as follows :  i) the DET regime is actually present  for $0<\alpha \leq 3$   while for a larger value $\alpha > 3$ the current decreases monotonically with the disorder strength $W$;
 ii) the DIT regimes disappears and seems to be a peculiar property of the infinite interaction range ($\alpha=0$);
 iii) the threshold at which the DET regime starts depends on the range of interaction $\alpha$ and on the system size;
 iv) from panel (b) we can infer that the DET regime is not a finite size effects since it persists for any $N$ considered.
In order to analyze in more details these results we divide our discussion into the cases $0<\alpha<1$ (strong long-range) and $\alpha \geq1$ (weak long-range and short-range). 

\section{Results and Discussion: strong long-range ($\alpha < 1$)}

In this section, we focus on identifying the critical disorder thresholds that define the boundaries of the Disorder-Enhanced Transport (DET) regime. Specifically, we derive $W_1^\alpha$, that marks the onset of the DET regime, and $W_{GAP}^\alpha$, that signals its end and that we identify as the value at which the local maximum of the current occurs. They provide a comprehensive framework to describe the transport properties as a function of disorder in the presence of distance-dependent interactions.
\subsection{Onset of DET : the threshold $W_1^\alpha$}
\begin{figure}[t]
\centering
\includegraphics[width = \columnwidth]{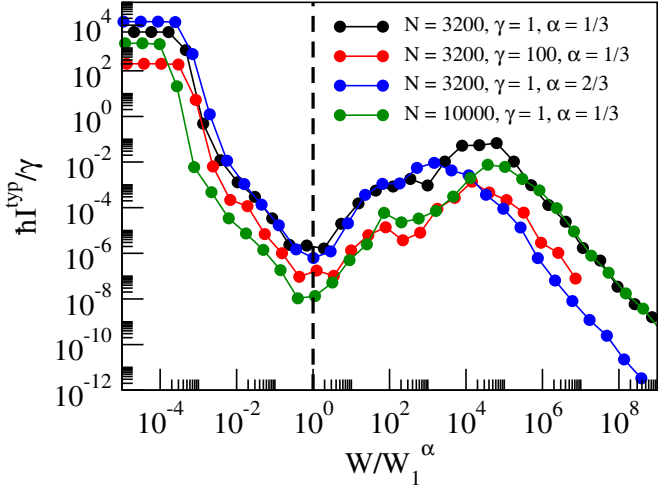}
\caption{Rescaled typical current $\hbar I^{typ}/\gamma$ as a function of the normalized disorder strength $W/W_1^\alpha$ for different system sizes $N$ and interaction parameters $\gamma$ and $\alpha$. Here, $\gamma_p=\gamma_d=1$ and $N\times N_r = 10^5$.
}
\label{fig:fig5}
\end{figure}

In our analysis of transport regimes, we observe that the transition to the Disorder-Enhanced Transport (DET) regime takes place at a well-defined disorder strength $W_1^\alpha$, which varies with the system size, the interaction amplitude $\gamma$ and the interaction range $\alpha$.\\
We found here that the disorder threshold which determines the DET regime is given by the phenomenological expression: 
\begin{equation}
    W_1^\alpha = \gamma(1-\frac{1}{2^\alpha}) \sqrt{{210.4\ln{N}}\over{N}}.
\label{W1alphath3}
\end{equation}
This equation can be   obtained by substituting into the formula   for the all-to-all case, see Eq.~(\ref{eq:DET}), an effective nearest-neighbor hopping $\Omega_{\alpha}$: 
\begin{equation}
\Omega_\alpha =  \gamma(1-\frac{1}{2^\alpha})
\label{effcoup}
\end{equation}
While the value of this effective hopping has been inspired by some considerations presented in Ref.~\cite{deng2018duality}, we do not have a full theoretical justification for this. In Appendices A and B we give some evidence for the existence of this effective nearest-neighbor hopping. 
Note that this effective nearest-neighbor hopping which decreases as the hopping range increases is consistent with the cooperative shielding effect present in long range hopping models~\cite{celardo2016shielding}.
\\
This formula successfully captures the scaling of the onset of DET observed across a wide range of parameters. To check the validity of our estimate, we show in Fig. \ref{fig:fig5}  the typical current $I^{typ}$   as a function of the rescaled disorder strength $W/W_1^{\alpha}$ for different sets of parameters $\alpha, N, \gamma$ as indicated in the legend.
In the panel the x-axis is rescaled by the expression of $W_1^\alpha$ in Eq. (\ref{W1alphath3}).
As one can see all rescaled curves present a unique local minimum at $W= W_1^\alpha$, confirming the consistency of the threshold.  More data supporting the validity of the scaling law can be found in Appendix E. Note that this threshold is valid also for $1< \alpha \leq 3$ as it will be discussed in the next sections.\\
\subsection{End of the DET regime and peak current at $W_{GAP}^\alpha$}
In order to derive a theoretical estimation for $W_{GAP}^\alpha$, we  follow the derivation of the Gap Equation \cite{jung1999resonance, cooper1956bound, schrieffer1964theory} proposed in \cite{chavez2019real} for the case of an all-to-all long-range hopping.
Let us consider the following Hamiltonian:
\begin{equation}
\hat{H} = \hat{H_0} + \hat{V}_\alpha = \sum_k E_k^0 \ket{k}\bra{k} - V_0\sum_{k,k'}{\ket{k}\bra{k'}\over{|k-k'|^\alpha}}
\end{equation}
where, in our case,  $E^0_k$ are the eigenvalues of $H_0$, and $\ket{k}$  are the correspondent eigenvectors (site basis).
For simplicity, let us assume for the latter a Picket-Fence distribution, namely
\begin{equation}
E^0_k = k\delta = k{W\over{N}}
\end{equation}
where $k = -{N\over{2}},...{N\over{2}}$ and $\delta = W/N$ is the level spacing.\\
In order to solve the Schr\"odinger Equation for the stationary states,
\begin{equation}
\hat{H}\ket{\Psi} = E\ket{\Psi},
\label{schr}
\end{equation}
we expand in site basis,
\begin{equation}
\ket{\Psi} = \sum_k a_k \ket{k}.
\label{expansion}
\end{equation}
Eq. (\ref{schr}) can be rewritten as 
\begin{equation}
(\hat{H}-\hat{H_0})\ket{\Psi} = \hat{V}_\alpha \ket{\Psi}
\end{equation}
Then, using  Eq.(\ref{expansion}), we get
\begin{equation}
(\hat{H}-\hat{H_0})\sum_{k'}a_{k'}\ket{k'} = \hat{V_\alpha} \sum_{k'}a_{k'}\ket{k'} = \sum_{k'}a_{k'}\hat{V_\alpha} \ket{k'}
\end{equation}
which leads to:
\begin{equation}
(\hat{H}-\hat{H_0})\sum_{k'}a_{k'}\ket{k'} = -\gamma\sum_{k'\ne k''}{a_{k'}\over{|k'-k''|^\alpha}}\ket{k''}
\label{schr1}
\end{equation}
Projecting  Eq.(\ref{schr1}) on the state $\bra{k}$ we get:
\begin{equation}
(E-E^0_k)a_k = -\gamma \sum_{k'\ne k}{a_{k'}\over{|k-k'|^\alpha}}
\end{equation}
and from that,
\begin{equation}
\sum_k a_k = -\gamma \sum_k {1\over{E-E_k^0}}\sum_{k'\ne k}{a_{k'}\over{|k-k'|^\alpha}}
\end{equation}
For large system size $N$, we can approximate the summations with integrals:
\begin{equation}
\int_{-N/2}^{N/2}D(k) dk = -\gamma\int_{-N/2}^{N/2} {dk\over{E-E^0_k}}\int_{-N/2}^{N/2} dk' \frac{D(k')}{|k-k'|^\alpha}
\end{equation}
where the coefficients $a_k$ have been replaced with the density of the levels which is independent of $k$, ($D(k)  = 1/\delta = N/W$).\\
By solving the integral on the left-hand-side and estimating the second intregral with its maximal value at $k =  0$ so that ${1\over{|k-k'|^\alpha}} \approx {1\over|k'|^\alpha}$, we can rewrite the last equation as
\begin{equation}
N \approx -2\gamma \int_{-N/2}^{N/2} {dk\over{E-E^0_k}} \int_{0}^{N/2}dk' (k')^{-\alpha}
\end{equation}
which leads to
\begin{equation}
N \simeq -2V_0{(N/2)^{1-\alpha}\over{1-\alpha}} \int_{-N/2}^{N/2} {dk\over{E-E^0_k}}
\end{equation}
Since  $E^0_k = k{W\over{N}}$ we can  solve  the  integral 
getting,
\begin{equation}
1 = 2V_0{(N/2)^{1-\alpha}-1\over{1-\alpha}} {1\over{W}} \ln{{2E-W}\over{2E+W}}
\end{equation}
Note that the above solution is valid only for $E<-W/2$.  
For large enough disorder (see Appendix C)  the first excited state is given by $-W/2$ so that the only state that satisfies this requirement is the ground state $E_1$, see also \cite{chavez2019real}.

Defining $C_\alpha \equiv {(N/2)^{1-\alpha}-1\over{1-\alpha}} $  the energy of the ground state 
is given by 
\begin{equation}
E_1 = {W\over{2}} {{1+\exp{({W\over{2\gamma C_\alpha}}})}\over{1-\exp{({W\over{2\gamma C_\alpha}}})}}
\end{equation}
By recalling that the energy of the first excited state $E_2 \rightarrow -W/2$ for sufficiently large   $W$ (see Appendix C), we have for the energy gap the following estimate:
\begin{equation}
\Delta^{th} = E_2 - E_1 = -{W\over{2}} - E_1 = {W\over{e^{W\over{2\gamma C_\alpha}}}-1}
\label{deltaPF}
\end{equation}
Let us note that when $W \gg \gamma$ we get
\begin{equation}
\Delta^{th} \simeq We^{-{W\over{2\gamma C_\alpha}}}
\end{equation}
By setting $\Delta = \delta$, it is easy to find that, for $N \gg 1$
\begin{equation}
W_{GAP}^\alpha \simeq 2\gamma {(N/2)^{1-\alpha}-1\over{1-\alpha}}\ln{N}
\label{Wgapth}
\end{equation}
Let us note that, when $\alpha = 0$, the definition of $W_{GAP}^\alpha$ matches the expression proposed in \cite{chavez2021disorder}. In addition, one can reverse the equation to obtain the critical value of the interactions strength $\gamma$ at which the energy gap opens:
\begin{equation}
\gamma_{cr} \simeq \frac{(1-\alpha)W}{2\ln{N}((N/2)^{1-\alpha}-1)}
\label{criticalgamma}
\end{equation}
\begin{figure}[t]
\centering
\includegraphics[width =\columnwidth]{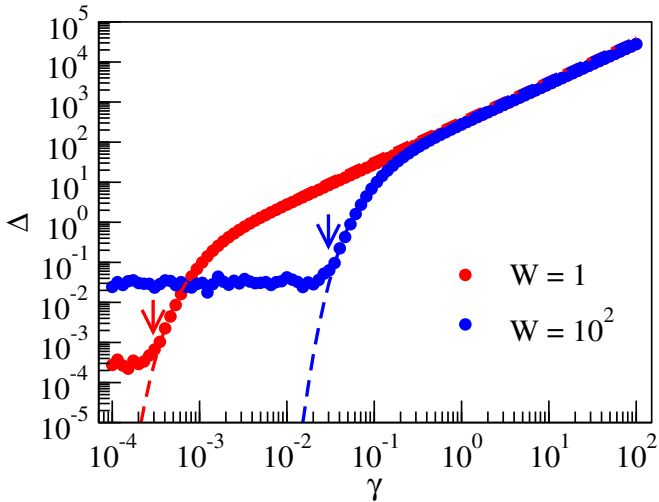}
\caption{Numerical energy gap $\Delta$ (dots) between the first excited state and the energy gap for $N = 3200$  and $W = 1$  and $W = 10^2$ as a function of the interaction strength $\gamma$. The dashed lines represent the theoretical prediction given in Eq. \ref{deltaPF}. The arrows indicate the interaction strength $\gamma_{cr}$ at which the energy gap opens given in Eq. (\ref{criticalgamma}).
Here, $\alpha = 1/3, \gamma_p=\gamma_d=1$ and $N\times N_r = 10^5$.}
\label{fig:fig6}
\end{figure}
A comparison between the $\Delta$ obtained using the modified Picket-Fence model and the exact energy gap calculated numerically as the energy difference between the first excited state and the ground state is shown in Fig. \ref{fig:fig6} for different disorder strengths $W$. As one can see, Eq.~(\ref{criticalgamma})  gives a good estimate of the disorder threshold where the gap opens, see arrows in Fig.~\ref{fig:fig6}, for the model given in Eq.~(\ref{ham}).
\begin{figure}[t]
\centering
\includegraphics[width = \columnwidth]{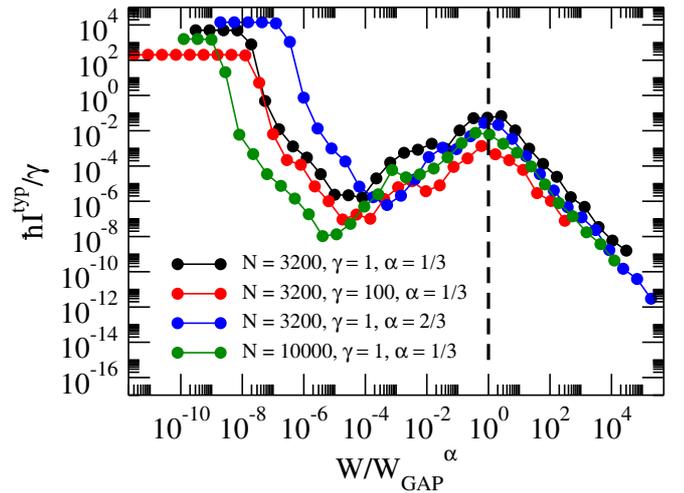}
\caption{Rescaled typical current $\hbar I^{typ}/\gamma$ as a function of the normalized disorder strength $W/W_{GAP}^\alpha$ for different system sizes $N$ and interaction parameters $\gamma$ and $\alpha$. Here, $\gamma_p=\gamma_d=1$ and $N\times N_r = 10^5$.}
\label{fig:fig7}
\end{figure}

In order to show the validity of our estimation of the local maxima of the current given by Eq.~(\ref{Wgapth}), 
in Fig. \ref{fig:fig7} the typical current $I^{typ}$ is shown as a function of the rescaled disorder strength $W/W_{GAP}^\alpha$ for several sets of parameters $N, \gamma$ and $\alpha$.   Since the positions of all the rescaled local maxima coincide we can assert the reliability of the equation for $W_{GAP}^{\alpha}$.
More data confirming the scaling law can be found in Appendix F.
\\

 \begin{figure*}[t]
\centering
\includegraphics[width=\textwidth]{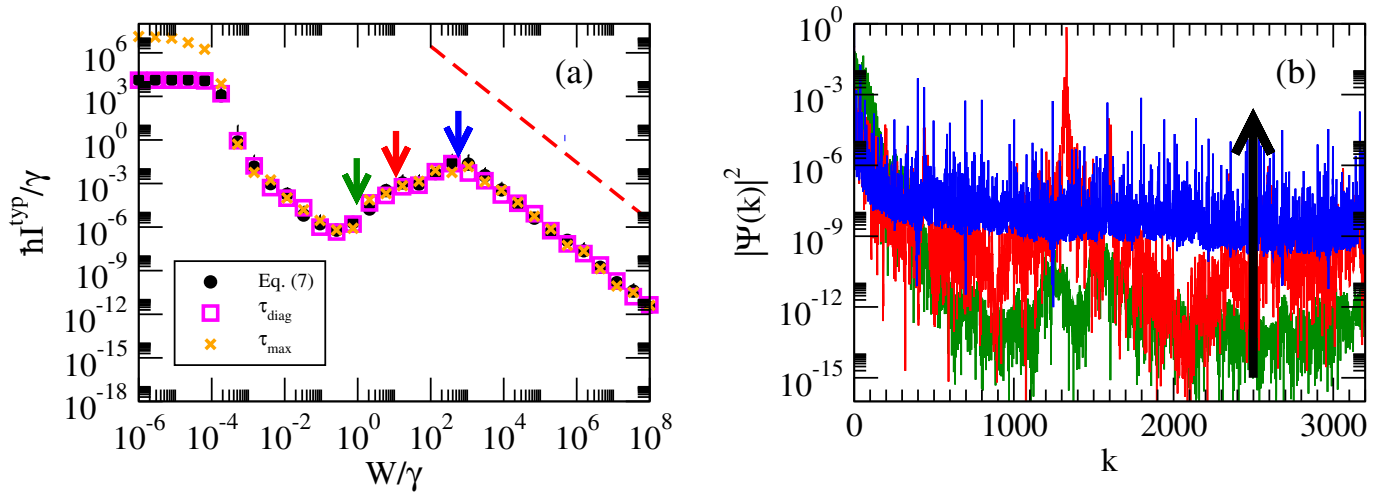}
\caption{
(a) Rescaled typical current $\hbar I^{typ}/\gamma$ as a function of disorder strength $W/\gamma$ for a chain of $N = 3200$ sites, a long-range hopping amplitude $\gamma = 1$ for an interaction range $\alpha = 2/3$ (black circles). Error bars are given by the standard deviation of the average over the different configurations.
Dashed red line indicates the behavior $1/W^2$. The pink squares represent the typical current computed with the simplied form in Eq. \ref{tau_k} while the orange cross represent the typical current computed only with the maximal $\tau_k$ among the elements of the sum of Eq. \ref{tau_k}.
(b) Probability $|\Psi(k)|^2$ for the most conducting state in the site basis $k$ for different disorder strengths in the DET regime. In (a) colored arrows indicate the corresponding disorder strengths on the typical current curve. In (b) a black arrow highlights the increasing amplitude of the tails as disorder grows. Here: $\alpha = 2/3, \gamma_p = \gamma_d = \gamma$ and $N \times N_r = 10^5$.
The probability $|\psi_k|^2$ for the most conducting state has been computed as follows: 
i) we consider the state that gives the maximum $\tau_k$ where $\tau_k$ are the elements of the sum in Eq. \ref{tau_k} and we repeat for several disorder configurations; ii) among the different states for different disorder configurations we select the one that gives the maximum typical current $I^{typ}$.}
\label{fig:fig4}
\end{figure*}
\subsection{The DET regime and the shape of eigenfunctions}

In Ref.~\cite{chavez2021disorder} the DET regime for $\alpha=0$ was explained by analyzing the shape of the eigenfunctions. Here we follow a similar path to explain the DET regime for generic long-range hopping. 

First of all, let us  observe that the typical current computed using the full expression for the transfer time $\tau$, as given in Eq.~(\ref{tau}), is numerically very close to the one obtained using a simplified formula where only the diagonal terms ($k = k'$) of the double sum are retained:
\begin{equation}
\tau_{diag} = \hbar \gamma_d \sum_k \frac{|\braket{N|r_k}|^2 |\braket{\tilde{r}_k|1}|^2 }{4\Gamma_k^2}
\label{tau_k}
\end{equation}
where $\Gamma_k$ are the imaginary parts of the complex eigenvalues $\epsilon_k$. Although we do not have a complete theoretical explanation for this agreement, we conjecture that it may be due to the interference effects between different eigenstates. In particular, destructive interference between off-diagonal terms could suppress their overall contribution to $\tau$, making the diagonal approximation accurate, as shown in Fig. \ref{fig:fig4}(a). Since  $\Gamma_k = \frac{\gamma_d}{2}|\braket{N|r_k}|^2$, it is easy to see that Eq.(\ref{tau_k}) can be written as,
\begin{equation}
\tau_{diag} = \frac{\hbar}{\gamma_d} \sum_k \frac{ |\braket{\tilde{r}_k|1}|^2 }{|\braket{N|r_k}|^2}
\label{taudiag}
\end{equation}
This expression reveals that the dominant contribution arises from the eigenstates that simultaneously maximize the ratio between the probability amplitude on the injection site ($\ket{1}$) and that on the extraction site ($\ket{N}$). This condition reflects a nontrivial interplay since an eigenstate that maximizes the nominator could be different from the eigenstate that minimizes the denominator. 

Digging a little bit more in Eq.~(\ref{taudiag}) one also find that for most disorder strengths (and surely in the DET region) the current is actually dominated only by the maximal element $\tau_{max}$ of the sum in  Eq. ~(\ref{taudiag}). In other words, {\it one single eigenstate} (the one with the largest overlap with the site where the excitation is pumped and the smallest with the site where the current is extracted) contributes to the total current. The current obtained with a single eigenstate is shown with orange crosses in Fig. \ref{fig:fig4}(a) and it should be compared with the total current (full black circles). 
 As can be seen, excluding a region characterized by a very small disorder strength $W \lesssim 10^{-4}$, there is excellent agreement with the total typical current. This confirms that in the Disorder-Enhanced Transport regime, the current  is almost uniquely determined by a single eigenstate.\\ 
To better understand the structure of this dominant state, we analyze its components $|\psi(k)|^2$ in the site basis in Fig. \ref{fig:fig4}(b). While the main peak remains localized near the injection site, the probability in the tails of the wavefunction increases with the disorder strength $W$ in the DET regime. This behavior provides a natural explanation for the persistence of the DET mechanism: the tails connecting input and output sites enhance transmission even at strong disorder, allowing the current to increase in a monotonic way with $W$.  

For $0<\alpha<1$ the tail of this dominant eigenstate always increases with the disorder strength,  up to the $W_{GAP}^{\alpha}$ and a regime where the tails are independent of the disorder strength is absent, at variance with the case $\alpha=0$ studied in Ref.~\cite{chavez2021disorder}. This is consistent with the absence of the DIT regime for finite hopping range. \\

\section{Extension to short-range interactions: the case $\alpha > 1$}
\begin{figure*}[t]
\centering
\includegraphics[width = \textwidth]{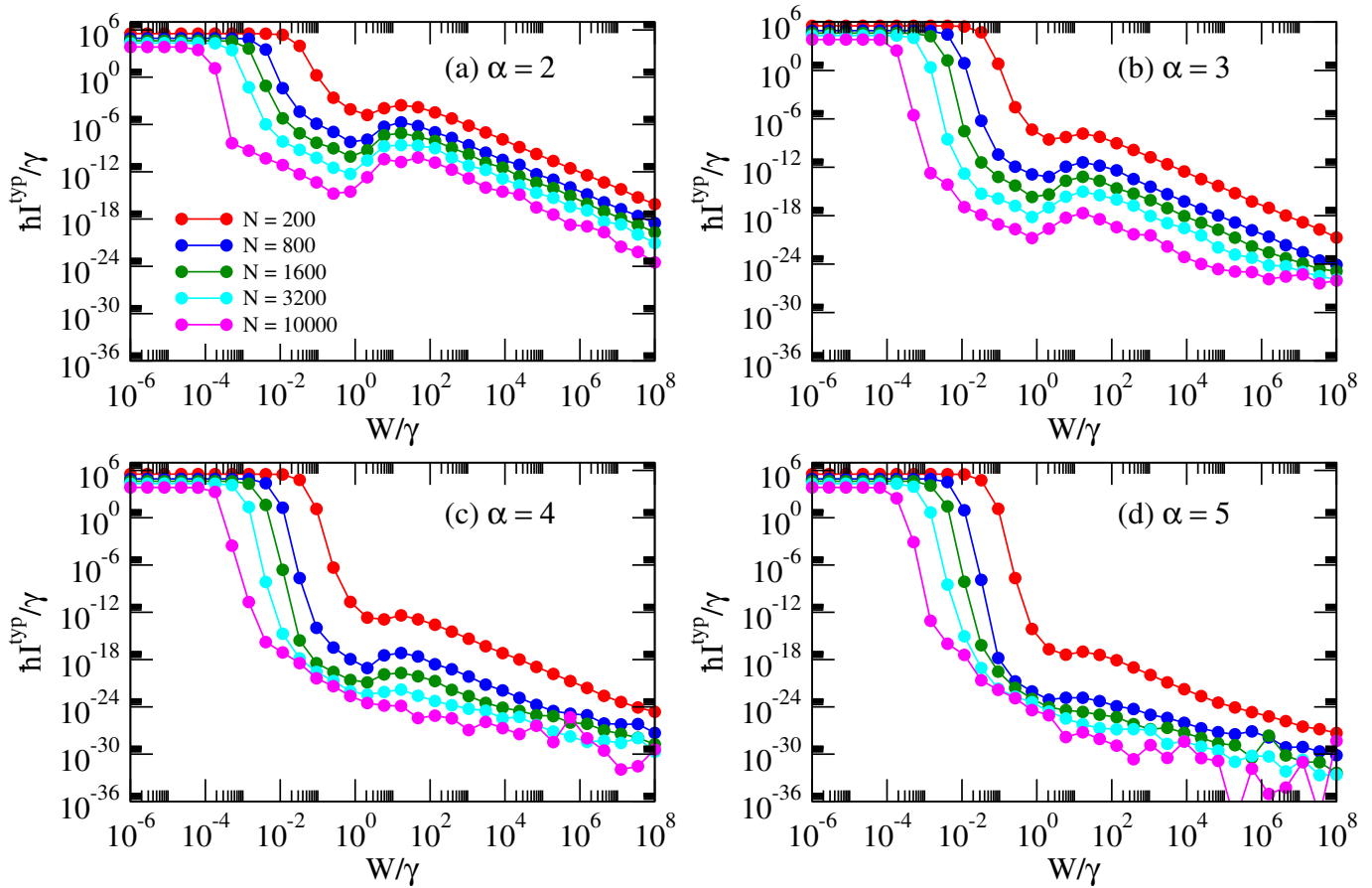}
\caption{Typical current $I^{typ}$ as a function of disorder strength $W$ for interaction exponents (a) $\alpha = 2$, (b) $\alpha = 3$, (c) $\alpha = 4$ and (d) $\alpha = 5$ and system sizes $N = 200, 800, 1600, 3200, 10000$. Here, $\gamma = \gamma_p=\gamma_d=1$ and $N\times N_r = 10^5$.}
\label{fig:sr}
\end{figure*}

In the analysis for $\alpha < 1$, we successfully estimated both the onset of Disorder-Enhanced Transport (DET), $W_1^\alpha$, and the critical disorder strength $W_{GAP}^\alpha$, which marks the local maximum of the current. 
To extend our analysis beyond the long-range interaction regime, we computed the steady-state currents also for $\alpha > 1$ using the same numerical approach described in the main text. When $\alpha > 1$, the energy gap between the extended ground state and the first excited state gradually closes \cite{celardo2016shielding}, leading to a progressive return to a localized transport regime. To investigate the persistence of the Disorder-Enhanced Transport (DET) regime for $\alpha > 1$, we calculated the typical current $I^{typ}$ for different values of $\alpha$ and system sizes $N$. The results are shown in Fig. \ref{fig:sr}. 

As can be seen, the DET peak remains visible also for  $\alpha =2,3$,  while for large system size $N$ it disappears when $\alpha=4,5$.  Even if we do not have any theoretical argument about  the critical $\alpha$ for which the DET regime persists in the thermodynamic limit, our numerical results seems to suggest that for   the DET regime is well defined   in the thermodynamic limit only when $\alpha \leq 3$. It is well known that in the region   $1<\alpha \leq 3$, also called weak long-range \cite{defenu2023long},  many properties of the strong long-range region persist. More theoretical investigations are necessary to estimate in a better way the threshold in $\alpha$ to observe the DET.

\begin{figure}[t]
\centering
\includegraphics[width =\columnwidth]{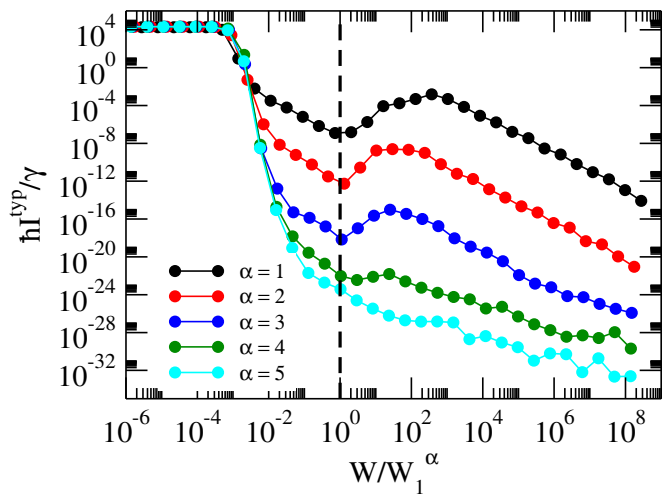}
\caption{Typical current $I^{typ}$ as a function of the normalized disorder strength $W/W_{1}^\alpha$ for $N = 3200, \gamma = 1$ and different short interaction ranges $\alpha$.  Here, $\gamma_p=\gamma_d=\gamma$ and $N\times N_r = 10^5$.}
\label{fig:fig9}
\end{figure}

The estimate   for the disorder threshold $W_1^\alpha$, see Eq. ~(\ref{W1alphath3}), which marks the beginning of the Disorder-Enhanced Transport regime, turns out to be valid also up to $\alpha=3$, as shown in Fig. \ref{fig:fig9}.
%

\begin{figure}[h]
\centering
\includegraphics[width = \columnwidth]{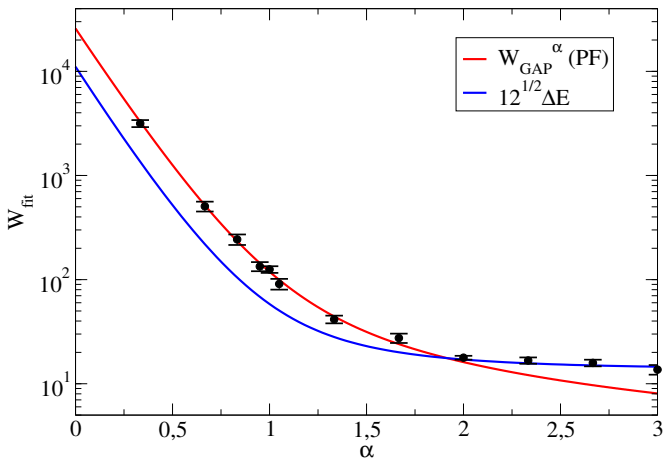}
\caption{Numerically extrapolated $W_{fit}$ (maximum of the parabolic fit of the typical current's local peak) as a function of the interaction parameter $\alpha$ for $N = 3200$ and $\gamma = 1$. Red lines represent the theoretical estimate in Eq. \ref{Wgapth} obtained using the Picket-Fence model while blue lines represent the numerically obtained spectral radius $\Delta E$ multiplied by $\sqrt{12}$.  Numerical values have been obtained by fitting the local peak with a parabolic function, and error bars are given by half the distance between the two numerical disorder strengths closest to the maximum of the fitting parabola.}
\label{fig:fig10}
\end{figure}

To obtain the critical disorder strength at which the DET ends,
we numerically fit the local maxima of the current with a parabolic function getting the value $W_{fit}$.  Results are shown in 
  Fig. \ref{fig:fig10} as black circles. As one can see the numerical points  follow the red line (i.e. $W_{GAP}^\alpha$) up to $\alpha \approx 2$.  
A simple rough criterion allows us to estimate the local maxima of the current even for $\alpha >2$,   by  setting the variance of the disorder equal  to the spectral radius $\Delta E$, namely when
\begin{equation}
\sqrt{\frac{W^2}{12}} \approx \Delta E,
\end{equation}
leading to 
\begin{equation}
W_{peak}^\alpha \approx \sqrt{12} \Delta E
\label{Wpeak}
\end{equation}
Results are shown in Fig. \ref{fig:fig10} as a blue curve. As one can see the agreement is excellent.
It is also possible (see Appendix D) to  obtain an analytical approximation for the spectral radius $\Delta E$:
\begin{equation}
\Delta E = 4\gamma \sum_{n=1}^{N/4}\frac{1}{(2n-1)^\alpha}
\end{equation}
This expression is    always finite for any $\alpha > 1$ and in the limit of large $N$ values it can be approximated as:
\begin{equation}
\Delta E \approx 4\gamma \zeta(\alpha)(1-\frac{1}{2^\alpha})
\end{equation}
where $\zeta(\alpha)$ is the Riemann zeta function. Therefore in the limit of $N\rightarrow \infty$ we obtain $W_{peak}^{\alpha} \approx 4\sqrt{12}\gamma\zeta(\alpha)(1-\frac{1}{2^\alpha})$.

\section{Conclusions}
Our analysis of a paradigmatic tight-binding one dimensional model with on-site disorder and tunable long-range hopping has revealed the emergence of a Disorder-Enhanced-Transport (DET) regime as a generic feature of long-range hopping systems. In the DET regime, contrary to the Anderson Localization paradigm, the current increases with the disorder strength, instead of being exponentially suppressed. 
We defined the disorder threshold where the DET regime begins as $W_1^{\alpha}$, Eq.~(\ref{W1alphath3}), and where the DET regime ends, in correspondence of a local maximum of the current, which is determined by the maximum between $W_{GAP}^\alpha$ and  $W_{peak}^{\alpha}$, Eqs~(\ref{Wgapth},\ref{Wpeak}).

In order to observe the DET regime it is necessary to have $W_{peak}^{\alpha}> W_1^{\alpha}$. 
For $\alpha<2$,  $W_{peak}^{\alpha}=W_{gap}^{\alpha}$ Eq.~(\ref{Wgapth}). Since $W_{gap}^{\alpha}$ increases with $N$ and $W_{1}^{\alpha}$ decreases with the system size, the DET regime will become more and more relevant as the system size increases. For  $2<\alpha<3$, $W_{peak}^{\alpha}$ is given by Eq.~(\ref{Wpeak}) and it is independent of the system size, while $W_{1}^{\alpha}$ decreases with the system size. Thus also in this case the DET regime occurs  for any system size.

Our results provide a comprehensive picture of disorder-assisted transport in systems with power-law interactions and suggest new routes to control energy flow in nanoscale devices. In particular, they are relevant to organic molecular aggregates \cite{schachenmayer2015cavity,orgiu2015conductivity}, Rydberg atomic ensembles \cite{jurcevic2014quasiparticle}, and other cavity QED architectures where long-range hopping naturally arises \cite{feist2015extraordinary}.

Altough the present work focuses on the single-excitation regime, it is known that long-range many-body systems exhibit quasi-invariant subspaces due to cooperative shielding \cite{santos2016cooperative}. This mechanism can effectively reduce the mixing of states induced by the interaction for certain initial configurations, suggesting that some of the transport features discussed here may persist beyond the one-particle limit. Even if a detailed analysis of the many-excitation regime lies beyond the scope of this work, it represents a promising direction for future investigations together with   the relevance of the DET regime in systems with larger dimension.

\section{Appendix A: Comparison with the case $\alpha = 0$}
\begin{figure*}[ht]
\centering
\includegraphics[width=\textwidth]{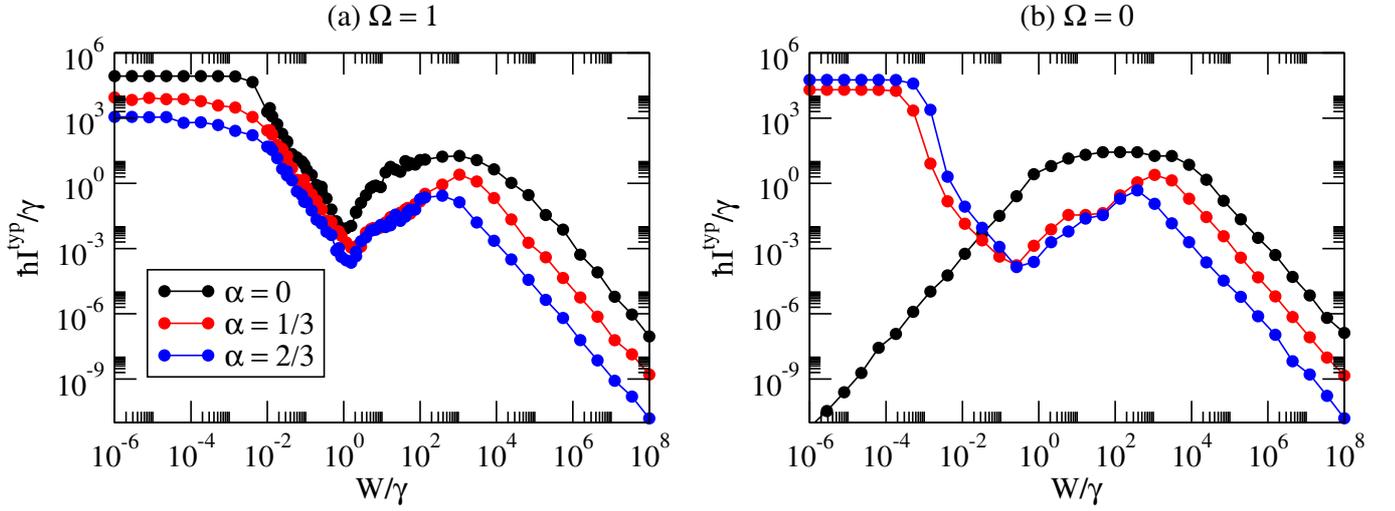}
\caption{(a) Typical current $I^{typ}$ as a function of disorder strength $W$ for a chain of $N = 800$ sites, a long-range hopping amplitude $\gamma = 1$, interaction ranges $\alpha = [0, 1/3, 2/3]$ and a nearest-neighbor hopping amplitude $\Omega = 1$. 
(b) Typical current $I^{typ}$ as a function of disorder strength $W$ for a chain of $N = 800$ sites, a long-range hopping amplitude $\gamma = 1$, interaction ranges $\alpha = [0, 1/3, 2/3]$ and a nearest-neighbor hopping amplitude $\Omega = 0$.
Here: $\gamma_p = \gamma_d = \gamma$  and $N \times N_r = 10^5$.
}
\label{fig:fig8}
\end{figure*}

In the model presented in \cite{chavez2021disorder}, in addition to the long-range hopping $\gamma$, a nearest neighbor coupling $\Omega$ was included, which plays a crucial role in determining transport properties. In particular, in the limit of all-to-all long-range interactions ($\alpha = 0$) and $\Omega \neq 0$, the typical current $I^{typ}$ presents a finite value in the absence of disorder. \\
Here, we explicitly set $\Omega = 0$ and consider a distance-dependent long-range interaction characterized by an interaction parameter $\alpha$. In the limit $\alpha = 0$ and $\Omega = 0$, the Hamiltonian becomes fully connected with equal long-range couplings $\gamma$ between all sites. The resulting hopping matrix is highly symmetric, leading to a trivial spectrum at $W =0$: the ground state is fully delocalized (a uniform superposition over all sites) with energy $-\gamma (N-1)$ while all other eigenstats are degenerate with energy $+\gamma$ and mutually orthogonal to the ground state. When a source and a drain are applied to the system (e.g., on site 1 and site $N$, respectively) by means of the use of the effective Hamiltonian of Eq. (\ref{effham}), most eigenstates remain unchanged since they acquire vanishing imaginary energy components $\Gamma_k$ (which correspond to vanishing overlap with the drain source since $\Gamma_k = \frac{\gamma_d}{2}|\braket{N|r_k}|^2$). Only two eigenstates have finite overlap with the drain site: (i) the fully symmetric ground state which acquires a width $\Gamma_{GS} = \frac{\gamma_d}{2N}$, thus vanishing when $N\rightarrow +\infty$ (ii) a previously degenerate eigenstate which acquires a width $\Gamma_k = \frac{\gamma_d}{2N}(N-1)$. Since we have seen that Eq. (\ref{taudiag}) well approximates the total transfer time and that the latter is proportional to the sum of the superpositions of the eigenstates  with the site 1 divided by their superpositions with the site $N$, we can easily understand that the eigenstates with $\Gamma_k = 0$ have an infinite transfer time  thus  dominating the total sum. When disorder is introduced, the symmetry of the eigenstates breaks down allowing them to acquire a finite width $\Gamma_k >0$. Thus, we get a   finite average time $\tau_k$ and  a net current through the system.\\
In contrast, our results demonstrate that for $\alpha \neq 0$, an effective nearest-neighbor coupling $\Omega_\alpha$ emerges, allowing the current to remain finite at low disorders even in the absence of $\Omega$. This difference is shown in Fig \ref{fig:fig8} where we show the typical current $I^{typ}$ as a function of the disorder strength $W$ for a chain with $\Omega = 1$ (left) and $\Omega = 0$ (right) respectively.
\section{Appendix B: Participation Ratio of the most conducting state}
\begin{figure}
\centering
\includegraphics[width = \columnwidth]{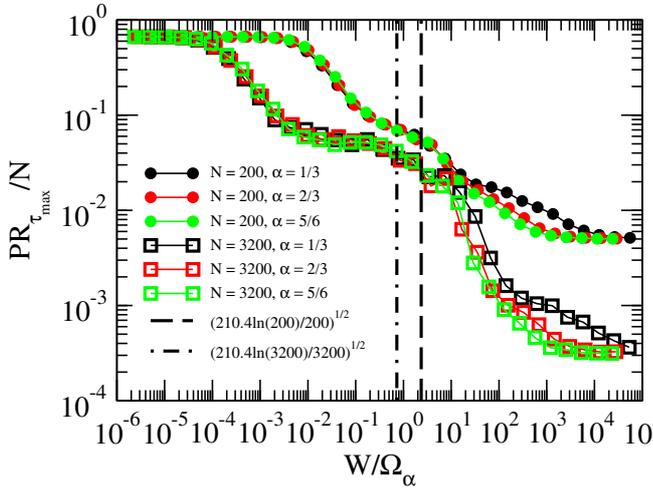}  
\caption{Normalized participation raio $PR_{\tau_{max}}/N$ of the eigenstate that gives the maximum contribution to the transport (i.e., the state associated with $\tau_{max}$) as a function of the rescaled disorder strength $W/\Omega_\alpha$, for different values of the interaction exponent $\alpha = 1/3, 2/3, 5/6$ and system sizes $N = 200, 3200$. The dashed (dot-dashed) line represents $W_1^\alpha/\Omega_\alpha$ for $N = 200$ ($N = 3200$). Here $\gamma = \gamma_d $ and we considered $N_r = 500$ disorder realizations for $N = 200$ and $N_r = 100$ realizations for $N = 3200$.}
\label{fig:figE1}
\end{figure}

In this section, we provide further support for the interpretation of the effective hopping amplitude $\Omega_{\alpha}$, introduced in Eq. (\ref{effcoup}), as the relevant energy scale that governs transport properties for a finite hopping range $\alpha$. Our goal is to show that $\Omega_\alpha$ plays a role analogous to the nearest-neighbor coupling $\Omega$ in the Anderson model, and that the dynamics in this regime is effectively controlled by it. To this end, we consider the eigenstate that gives the dominant contribution to transport, that is, the eigenstate associated with the maximum transfer time $\tau_k$ in Eq. (\ref{tau_k}), which we refer to as the \textit{most conducting eigenstate}. As discussed in the main text, when the disorder increases after the initial plateau, the current is largely dominated by a single eigenstate, whose projection on the site basis encodes the key transport features.\\
We analyze the degree of localization of this eigenstate $| \psi\rangle $ by computing its Participation Ratio, defined as:
\begin{equation}
PR = \frac{1}{\sum_{j=1}^{N}|\langle j |\psi\rangle |^4}.
\label{pr}
\end{equation}
We highlight that for a fully extended state, $PR = N$, while for a single-site localized state $PR  = 1$. We compute the $PR$ of the most conducting eigenstate state for $N = 200$ and $N = 3200$ and different values of the interaction exponent $\alpha$, and plot it as a function of the disorder strength $W$. To isolate the role of the effective coupling, we normalize the disorder strength by the effective hopping $\Omega_\alpha$ defined in Eq. (\ref{effcoup}), and the $PR$ by the system size $N$. The resulting curves $PR_{\tau_{max}}/N$ versus $W/\Omega_\alpha$ are shown in Fig. \ref{fig:figE1}.
The most striking feature of this plot is the collapse of the curves for different $\alpha$ onto a single one depending on $N$. This behavior strongly indicates that the localization properties of the dominant eigenstate depend solely on the ratio $W/\Omega_\alpha$ and on $N$, and not on the individual values of $W$ and $\alpha$. This results provide a   justification for the use of $\Omega_\alpha$ as the relevant energy scale in the system: it confirms that the dominant eigenstate "feels" the effective coupling $\Omega_\alpha$ as if it were a true nearest-neighbor hopping amplitude.

\section{Appendix C: Justification of $E_2 \rightarrow - \frac{W}{2}$}
\begin{figure}
\centering
\includegraphics[width = \columnwidth]{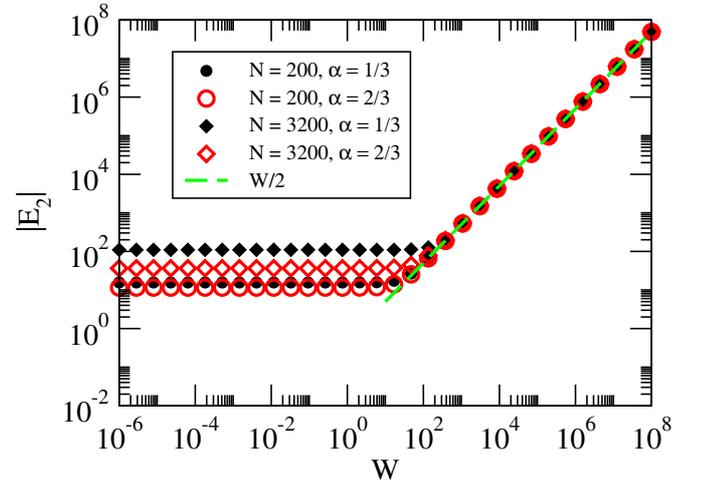}
\caption{Absolute value of the first excited state energy $E_2$ as a function of the disorder strength $W$ for $N = 200, 3200$ (respectively shown as circles and diamonds) and $\alpha = 1/3, 2/3$ (respectively in black and in red). The green dashed line represents $W/2$. Here: $\gamma = 1$, $N \times N_r = 10^5.$}
\label{fig:figC1}
\end{figure}

In Eq. (\ref{deltaPF}) we approximate
 the first excited state of the Picket-Fence model $E_2$ with the value    $\frac{W}{2}$ when  deriving  of the energy gap $\Delta$. It is easy to    analyze the behavior of $E_2$ as a function of the strength of the disorder $W$. In Fig. \ref{fig:figC1} we show the behavior of the energy of the first excited state (in absolute value) as a function of $W$ for different $\alpha$ and system size $N$.

As one can see the value of $E_2$ is approximately constant up to some small disorder strength  
 after which it becomes   $\propto W/2$ (see green dashed line in the same figure) as claimed in the text.
\label{Wint}

\section{Appendix D: Spectral radius for $\alpha > 1$}
\begin{figure}
\centering
\includegraphics[width = \columnwidth]{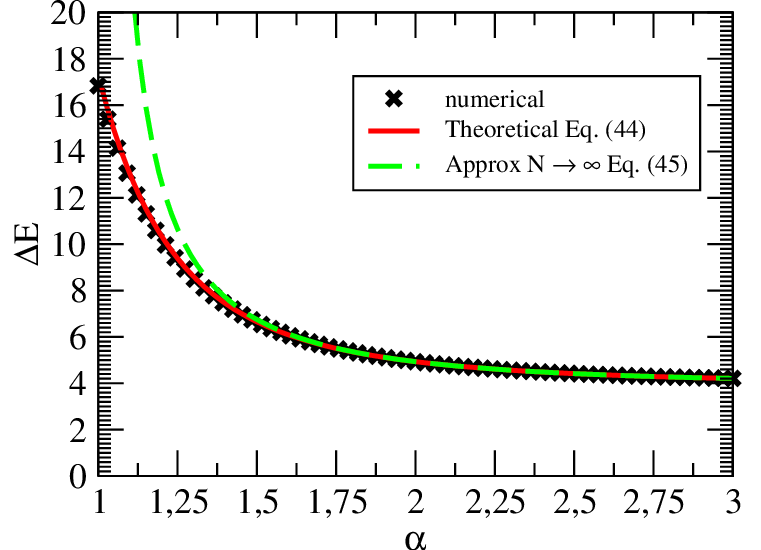}  
\caption{Spectral radius $\Delta E$ as a function of the interaction range $\alpha$ for a fixed system size $N = 3200$. Dots are numerical data, the full line is the analytical approximation see Eq. \ref{spectralrad} and the dashed line represents the $N\rightarrow \infty$ approximation of Eq. \ref{deltaEapprox}.
}
\label{fig:fig_spectralradius}
\end{figure}

Let us now compute the spectral radius of the Hamiltonian $H$ of Eq. (\ref{ham}) in the absence of disorder. In order to simplify the treatment we consider periodic boundary conditions. In such a case, the eigenvalues of $V_\alpha$ are (see \cite{celardo2016shielding}):
\begin{equation}
    E_q^{LR}=\left\{ 
    \begin{split}
        -2\gamma \sum_{n=1}^{N/2-1} \frac{\cos{2\pi q n/N}}{n^\alpha} \quad  N \quad  \text{Odd}\\
        -\gamma \frac{(-1)^q}{(N/2)^\alpha}-2\gamma \sum_{n=1}^{N/2-1} \frac{\cos{2\pi q n/N}}{n^\alpha} \quad  N \quad \text{Even}
    \end{split}\right.
\label{eigenvaluesalpha}
\end{equation}
Let us now suppose that $N$ is a multiple of 4. It is easy to see that the the maximum eigenvalue corresponds to $q = N/2$ while the minimum one corresponds to $q = N$:
\begin{equation}
\begin{split}
E_{\text{max}} = E_{N/2} &= -\gamma \frac{(-1)^{N/2}}{(N/2)^\alpha}
- 2\gamma \sum_{n=1}^{N/2 - 1} \frac{\cos{(\pi n)}}{n^\alpha} \\
&= -\gamma \frac{(-1)^{N/2}}{(N/2)^\alpha}
- 2\gamma \sum_{n=1}^{N/2 - 1} \frac{(-1)^n}{n^\alpha} \\
E_{\text{min}} = E_{N} &= -\gamma \frac{(-1)^N}{(N/2)^\alpha}
- 2\gamma \sum_{n=1}^{N/2 - 1} \frac{\cos{(2\pi n)}}{n^\alpha} \\
&= -\gamma \frac{(-1)^{N/2}}{(N/2)^\alpha}
- 2\gamma \sum_{n=1}^{N/2 - 1} \frac{1}{n^\alpha}
\end{split}
\end{equation}
This leads to:
\begin{equation}
\begin{split}
\Delta E &= E_{\text{max}} - E_{\text{min}} 
= 2\gamma \sum_{n=1}^{N/2 - 1} \frac{1 - (-1)^n}{n^\alpha} \\
&= 4\gamma \sum_{n=1}^{N/4} \frac{1}{(2n - 1)^\alpha}
\end{split}
\label{spectralrad}
\end{equation}
Eq. (\ref{spectralrad}) can be further approximated in the large $N$ limit giving 
\begin{equation}
\Delta E \approx 4\gamma\zeta(\alpha)(1-\frac{1}{2^\alpha})
\label{deltaEapprox}
\end{equation}
The comparison between the numerically computed spectral radius and the Eqs. (\ref{spectralrad}, \ref{deltaEapprox}) is given in Fig. \ref{fig:fig_spectralradius} \\
As one can see, the agreement between the numerical data obtained with our model and the model with the periodic boundary conditions is very good. The approximation of $N\rightarrow \infty$ works only in the region $\alpha \gtrsim 2$ because for $\alpha < 2$ finite $N$ corrections are quite strong. In any case, the spectral radius is finite for any $1<\alpha<3$ giving rise to a $W_{peak}^\alpha$ also finite in the thermodynamic limit $N\rightarrow \infty$.

\section{Appendix E: $W_1^\alpha$ scaling law}
\begin{figure*}
\centering
\includegraphics[width = \textwidth]{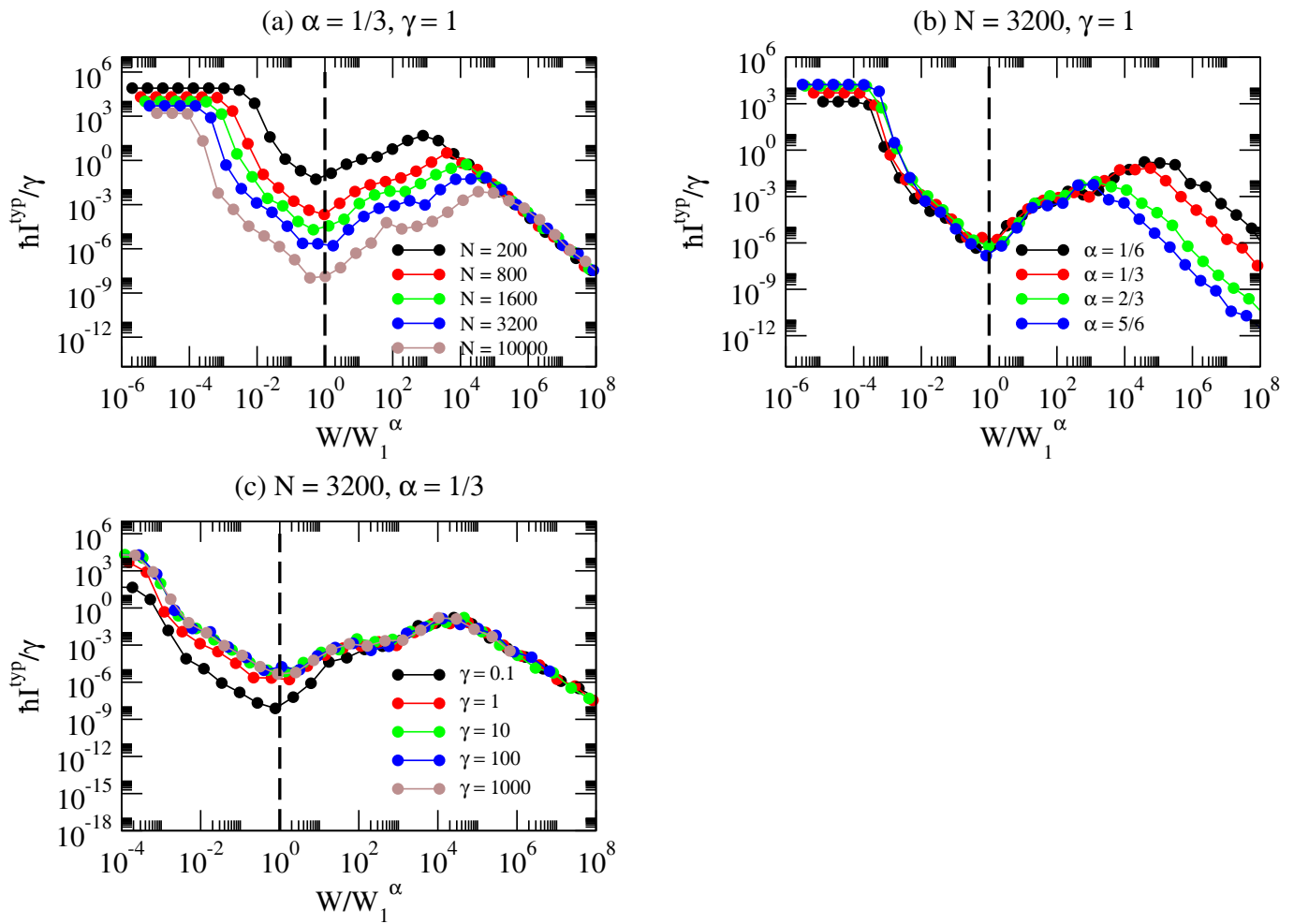}
\caption{(a) Rescaled typical current $\hbar I^{typ}/\gamma$ as a function of the normalized disorder strength $W/W_1^\alpha$ for a fixed interaction range $\alpha = 1/3$ and a fixed interaction strength $\gamma = 1$ for different system sizes $N$. (b) Rescaled typical current $\hbar I^{typ}/\gamma$ as a function of the normalized disorder strength $W/W_1^\alpha$ for a fixed system size $N$ and a fixed interaction strength $\gamma = 1$ for different interaction ranges $\alpha$. (c) Rescaled typical current $\hbar I^{typ}/\gamma$ as a function of the normalized disorder strength $W/W_1^\alpha$ for a fixed system size $N$ and a fixed interaction range $\alpha = 1/3$ for different interaction strengths $\gamma$. Here, $\gamma = \gamma_p=\gamma_d=1$ and $N\times N_r = 10^5$.}
\label{fig:supinfoE}
\end{figure*}

In this section we present additional results confirming the scaling law for the onset of the DET regime given in Eq. \ref{W1alphath3}. In Fig. \ref{fig:supinfoE}(a) we vary the system size at fixed interaction range, in Fig. \ref{fig:supinfoE}(b) we explore different interaction exponents $\alpha$ and in Fig. \ref{fig:supinfoE}(c) we change the hopping amplitude $\gamma$ at fixed size and interaction range. In all cases, the scaling law in Eq. (\ref{W1alphath3}) provides a robust estimate for $W_1^\alpha$.

\section{Appendix F: $W_{GAP}^\alpha$ scaling law}
\begin{figure*}
\centering
\includegraphics[width = \textwidth]{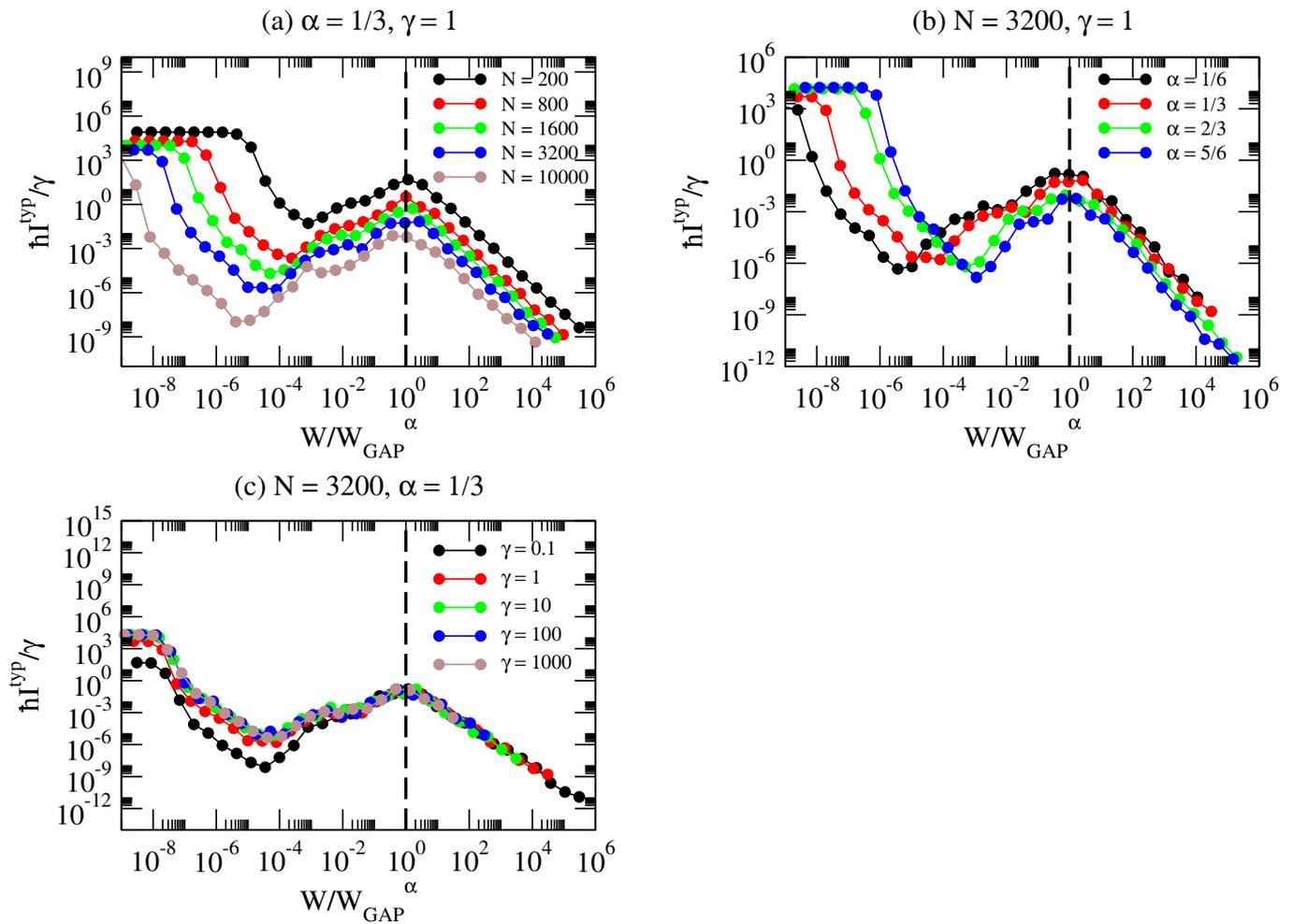}
\caption{(a) Rescaled typical current $\hbar I^{typ}/\gamma$ as a function of the normalized disorder strength $W/W_{GAP}^\alpha$ for a fixed interaction range $\alpha = 1/3$ and a fixed interaction strength $\gamma = 1$ for different system sizes $N$. (b) Rescaled typical current $\hbar I^{typ}/\gamma$ as a function of the normalized disorder strength $W/W_{GAP}^\alpha$ for a fixed system size $N$ and a fixed interaction strength $\gamma = 1$ for different interaction ranges $\alpha$. (c) Rescaled typical current $\hbar I^{typ}/\gamma$ as a function of the normalized disorder strength $W/W_{GAP}^\alpha$ for a fixed system size $N$ and a fixed interaction range $\alpha = 1/3$ for different interaction strengths $\gamma$. Here, $\gamma = \gamma_p=\gamma_d=1$ and $N\times N_r = 10^5$.}
\label{fig:supinfoF}
\end{figure*}

In this section we provide further evidence for the validity of Eq. (\ref{Wgapth}), which determines the end of the DET regime. In Fig. \ref{fig:supinfoF}(a) we compare different system sizes, in Fig. \ref{fig:supinfoF}(b) we vary the interaction exponent $\alpha$ and in Fig. \ref{fig:supinfoF}(c) we change the hopping strength $\gamma$ keeping fixed the other parameters. The position of the maxima consistently follow the predicted threshold $W_{GAP}^\alpha$, supporting the universality of the scaling law.

\clearpage
\newpage
\bibliographystyle{apsrev4-2}

\end{document}